# Controlling the polarization of nitrogen ion lasing

Jingsong Gao[1], Xiang Zhang[2], Yang Wang[1], Jiahao Dong[2], Mingwei Lei[1], Yi Liu[2,3,†], Chengyin Wu[1], Qihuang Gong[1] and Hongbing Jiang[1,‡]

[1]*State Key Laboratory for Mesoscopic Physics, School of Physics, Peking University, Beijing 100871, China*

[2]*Shanghai Key Lab of Modern Optical System, University of Shanghai for Science and Technology, Shanghai 200093, China*

[3]*CAS Center for Excellence in Ultra-intense Laser Science, Shanghai, 201800, China*

† yi.liu@usst.edu.cn
‡ hbjiang@pku.edu.cn

## Abstract

Air lasing provides a promising technique to remotely produce coherent radiation in the atmosphere and attracts continuous attention. However, the polarization properties of $N_2^+$ lasing with seeding have not been understood since it was discovered ten years ago, in which the polarization behaviors appear disordered and confusing. Here, we performed an experimental and theoretical investigation on the polarization properties of $N_2^+$ lasing and successfully revealed its underlying physical mechanism. We found that the optical gain is anisotropic owing to the permanent alignment of $N_2^+$ induced by the preferential ionization of the pump light. As a result, the polarization of the $N_2^+$ lasing tends to align with that of the pump light after amplification, which becomes more pronounced as the amplification factor increases. Based on the permanent alignment of $N_2^+$, we built a theoretical model that analytically interpreted and numerically reproduced the experimental observations, which points out the key factors for controlling the polarization of $N_2^+$ lasing.

The cavity-free lasing action of air plasma excited by high power femtosecond pulses gives rise to coherent optical radiations in backward directions [1], which has unique potential applications compared with traditional fluorescence detection [2] in atmospheric remote sensing. In addition to its capacity for remote detection of trace gases and isotopes, very recently, the lasing of nitrogen ions has also been exploited to generate ultraviolet supercontinuum and structured light, uncovering some new physical mechanisms [3-5]. Among the various lasing effects of nitrogen, oxygen, as well as argon gas, the coherent narrow-bandwidth emissions of $N_2^+$ around 391 nm, corresponding to the transition of $B^2\Sigma_u^+(v''=0) \rightarrow X^2\Sigma_g^+(v=0)$, have attracted extensive investigations [6-22]. In the case of the pump pulse with a central wavelength of 800 nm, the presence of the optical gain has been widely confirmed by the observation of strong amplification of an external seed pulse, and the $N_2^+$ coherent emission was considered as superfluorescence (SF) [6,13,14]. However, the underlying physical mechanism for optical gain around 391 nm has been intensively debated in the community [6,8,9,12,18,21,22], and the arguments for population inversion continue to this day [23,24].

Besides the controversy over the gain mechanism, one fundamental puzzle is related to the polarization property of the amplified SF at 391 nm with the injection of a seeding pulse. Hitherto, no model has been proposed to characterize and quantify its polarization behaviors since the nitrogen ion air lasing was discovered ten years ago. In particular, when the polarizations of the linearly polarized seed and pump pulses are perpendicular or parallel, the polarization of the amplified SF is identical to that of the seed light [25,26]. In contrast, when the polarizations of the seed and pump light form an angle, the polarization of the amplified SF is different from those of both the seed and pump light, which seems disordered and is unexpected for a classical laser amplifier. In the case of vector pump light [4], the amplified SF exhibits the same vector polarization characteristics as the pump light regardless of the seed light's polarization state, which further complicates the understanding of polarization properties in $N_2^+$ lasing.

H. Xie *et al*. explained why the polarizations of the P and R branches of 391 nm lasing are mirror-symmetrical about the polarization of the pump light [26], where the coupling of the *X* and *A* states driven by the 800 nm pump pulse leads to the negative coherence between the rotation states of *J* and *J*+2 in the *X* state. However, the underlying mechanism for the fact that the polarization properties of the $N_2^+$ SF exhibit disordered behaviors under different conditions remains unexplored. Polarization is one of the fundamental properties of light field, and its yet unclear mechanism manifests as a main challenge for the complete understanding and control of nitrogen ion air lasing action. As a promising remote coherent light source, comprehensive manipulation of its polarization state is now exceedingly necessary and urgent to customize nitrogen ions lasing toward higher-dimensional structured light, which has currently sparked tremendous interest across various fields [27-31].

In this work, we studied the polarization properties of $N_2^+$ lasing and efficiently revealed the underlying physical mechanism. We found the nitrogen ions were permanently aligned along the pump light polarization, which was characterized by the anisotropic polarization angular distribution of the lateral fluorescence of $N_2^+$. Due to the transition dipole moment between $X$ and $B$ states being parallel to the $N_2^+$ molecular axis, the optical gain exhibits anisotropy, with the maximum in the direction of pump light's polarization. Therefore, the polarization of $N_2^+$ lasing rotates towards that of the pump light with the amplification factor of the seed light increasing. Through a strict derivation based on the permanent alignment of $N_2^+$, we built a theoretical model that well reproduced the experimental results. The model describes an anisotropic gain medium and establishes the physical connection among the polarization directions of the $N_2^+$ lasing, the seed pulse and the pump pulse, pointing out the key factors for quantitatively controlling the polarization of $N_2^+$ lasing emission.

The schematic of the experimental setup is depicted in Fig. 1(a). The *y*-polarized femtosecond laser pulses (800 nm, 35 fs, 1.9 mJ) served as the pump beam. When the pump light was focused in 10 mbar nitrogen gas by a lens with a focal length of 30 cm (L1), a laser plasma filament emerged. A polarizer (P) was utilized to analyze the polarization characteristic of the lateral fluorescence. Fig. 1(b) shows that the measured spectral intensity of *y*-polarized fluorescence at 391 nm is stronger than that of *z*-polarized fluorescence. We then rotated the axis of the polarizer, varying its angle relative to the *y* direction, $\theta_y$. We obtained the angular distribution of lateral fluorescence at 391 nm as a function of $\theta_y$, as shown in Fig. 1(c). Because the electronic transition moment of *B* and *X* states is parallel to the molecular axis of $N_2^+$, the anisotropic angular distribution of lateral fluorescence indicates that the nitrogen ions in the *B* state were macroscopically aligned along the pump light's polarization. Previous research found that the lifetime of $N_2^+$ fluorescence is approximately hundreds of picoseconds at pressures of several millibars [32]. The long-life fluorescence exhibits an anisotropic angular distribution relative to the pump light polarization, suggesting the existence of the permanent (time-averaged) alignment of $N_2^+$.

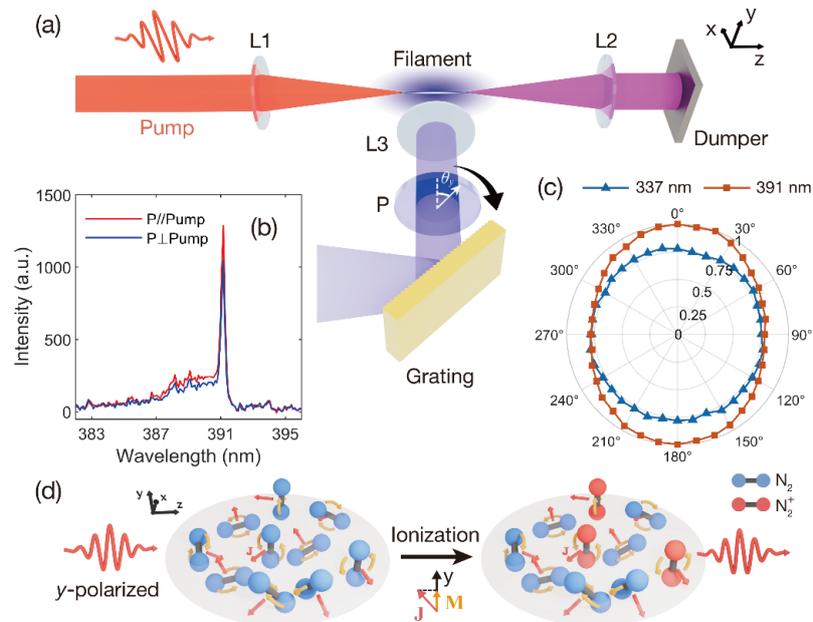

Fig. 1 (a) Schematic of the experimental setup for collecting lateral fluorescence. L1, L2, L3, lenses; P, polarizer. $\theta_y$ is the included angle between the polarization direction of the pump light and the polarization axis of P, namely, the angle relative to the *y* axis (b) Spectra of lateral fluorescence of $N_2^+$ at 391 nm. Shown are measured spectra when the polarization axis of P is perpendicular (black line) and parallel (red line) to the polarization of the pump light, respectively. (c) Angular distributions of lateral fluorescence of $N_2^+$ (391 nm) and $N_2$ (337 nm) as a function of $\theta_y$. (d) Illustration of preferential ionization of nitrogen molecules. J represents the angular momentum of a nitrogen molecule, M is the projection of J on *y* direction.

The *X* and *B* states are ionized from the highest occupied molecular orbital (HOMO) and the second lower-lying orbital (HOMO-2) of nitrogen molecules, respectively. According to MO-ADK theory [33], the optimal ionization directions of these two orbitals are parallel to the nitrogen molecular axis. As the angle between the electric field and the molecular axis increases, the ionization yield decreases, reaching a minimum when the electric field is perpendicular to the molecular axis. In Fig. 1(d), a schematic diagram illustrating the preferential ionization process of $N_2$ for *X* and *B* states, in which the nitrogen molecules with molecular axes parallel to the electric field are selected to be ionized. In the *JM*-representation, the molecules with higher probability of being ionized have a smaller helicity quantum number *M*, resulting in significant permanent molecular alignment of $N_2^+$ along the pump light's polarization.

As a simple illustration in Fig. 1(d), the ionized nitrogen molecules have a quantum number of $M = 0$. We measured the polarization angular distribution of the lateral fluorescence of neutral $N_2$ at 337 nm to be isotropic, as shown in Fig. 1(c). This indicates that the preferential ionization serves as the primary contributor to the permanent alignment of $N_2^+$.

Because the polarization of the pump light is perpendicular to the $xz$ plane, the average alignment degree in the $x$ direction is equivalent to that in the $z$ direction. Consequently, an $y$-polarized seed beam is expected to have larger amplification factor than a $x$-polarized one. We study this by injecting linearly polarized seed pulses with different polarization directions and intensities, as depicted in Fig. 2(a). A polarizer (P1) was used to control the angle between the polarization directions of the pump and seed light, $\alpha$. A half waveplate (HWP) was applied to adjust the seed energy. At first, the polarization of the seed light was successively set to be $y$-polarized ($\alpha = 0°$) and $x$-polarized ($\alpha = 90°$). The corresponding spectra of the $N_2^+$ SF at 391 nm with week (solid lines) and strong (dash lines) seed energies are shown in Fig. 2(b). The experimental results indicate that the amplification factor in the $y$ direction is always higher than that in the $x$ direction, which is consistent with the existence of the $y$-direction permanent alignment of $N_2^+$. In these cases ($\alpha$ is equal to 0° or 90°), the polarization directions of SF are all identical to that of the seed light.

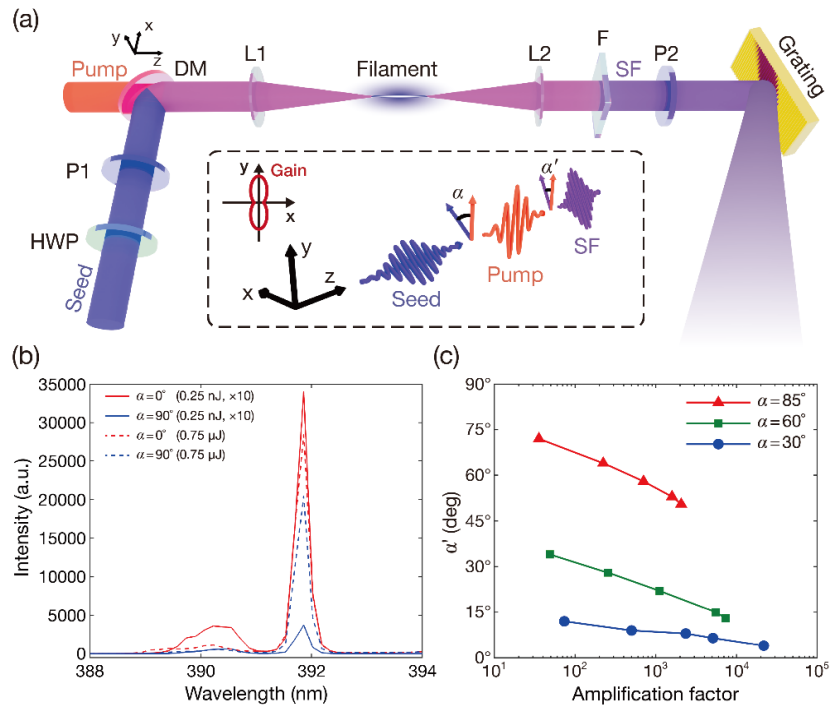

Fig. 2 (a) Schematic of the pump-seed setup. HWP, half-wave plate; P1, P2, polarizers; DM, dichroic mirror; L1, L2, lenses; F, filter. Inset: the red, blue and purple arrows indicate the polarization directions of the pump light, seed light and SF, respectively; $\alpha$ represents the angle between the polarization directions of the pump and seed light; $\alpha'$ represents the angle between the polarization directions of the pump light and SF. (b) Spectra of the amplified SF at 391 nm. The seed light is set weak, 0.25 nJ and strong, 0.75 μJ. (c) Measured $\alpha'$ as a function of amplification factor.

Comparing the results of the weak and strong seed pulses in Fig. 2(b), we noticed that the amplification ratio of the *y*-direction to the *x*-direction changes with the seed energy, the corresponding ratios are about 9.1 and 1.4, respectively. Due to the limited gain of $N_2^+$ medium, seed light with different energies corresponds to different amplification factors. This suggests that the polarization of the amplified SF will change with the amplification factor if seed electric field has both *x* and *y* components ($\alpha$ is unequal to 0° or 90°). Thus, we experimentally varied the seed energy to change the amplification factor and recorded the angle between the polarization directions of the pump light and SF, $\alpha'$, as shown in Fig. 2(c). We present the results of the P branch because its narrow bandwidth makes it easy to identify the polarization direction, while the case of the R branch is slightly different and will be discussed later. It is evident that the polarization direction of the $N_2^+$ lasing lies between that of the pump and seed light. Regardless of the value of $\alpha$ (except 0° and 90°), the SF polarization approaches the pump light's polarization as the amplification factor of the seed light increases.

To understand the underlying physical relationship between the polarization direction of $N_2^+$ air lasing and the amplification factor of the seed light, we developed a concise theoretical model. Here, we focused on how the spatial anisotropy of $N_2^+$ gain medium affects the polarization of the amplified seed light, neglecting the time-dependent aspect. When a nitrogen ion is stimulated by the seed light field, the slowly varying part of the corresponding induced dipole moment can be described by [34]

$$\mathbf{p} = -\frac{i}{\gamma}\Omega_R \mathbf{d}_{BX}, \qquad (1)$$

where $\gamma$ is the decay rate, $\Omega_R$ represents the Rabi frequency equal to $\mathbf{d}_{BX} \cdot \mathbf{E}/\hbar$, $\mathbf{d}_{BX}$ is the transition dipole moment. As mentioned above, the electronic transition moment of $B^2\Sigma_u^+(v"=0) \rightarrow X^2\Sigma_g^+(v=0)$ is parallel to the molecular axis of $N_2^+$. Therefore, $\mathbf{d}_{BX}$ is parallel to the molecular axis and can be expressed as $\mathbf{d}_{BX} = d_{BX}\mathbf{e}_m$, where $d_{BX}$ is the amplitude of the transition dipole moment, and $\mathbf{e}_m$ is the unit vector parallel to the molecular axis of $N_2^+$. The molecular vector can be expressed as

$\mathbf{e}_m = \cos\theta_x \mathbf{e}_x + \cos\theta_y \mathbf{e}_y + \cos\theta_z \mathbf{e}_z$, where $\mathbf{e}_x$, $\mathbf{e}_y$ and $\mathbf{e}_z$ are the unit vectors parallel to the *x, y* and *z* axes, respectively, $\theta_x$, $\theta_y$ and $\theta_z$ are the angles relative to *x, y*, and *z* axes, respectively. The expression of the seed light is written as $\mathbf{E} = E_x \mathbf{e}_x + E_y \mathbf{e}_y$, where $E_x$ and $E_y$ represent the *x* and *y* components of the electric field, respectively. Substituting these expressions into Eq. (1), we have

$$\mathbf{p} = -\frac{i}{\gamma\hbar} d_{BX}^2 [(E_x \cos^2\theta_x + E_y \cos\theta_x \cos\theta_y)\mathbf{e}_x + (E_x \cos\theta_x \cos\theta_y + E_y \cos^2\theta_y)\mathbf{e}_y]. \quad (2)$$

The gain medium consists of numerous nitrogen ions. The slowly varying component of the macroscopic polarization of the $N_2^+$ medium (dipole moment per unit volume) is given by

$$\mathbf{P} = N\langle\mathbf{p}\rangle, \quad (3)$$

where *N* represents the inversion number of nitrogen ions per unit volume, $\langle\mathbf{p}\rangle$ is the average of $\mathbf{p}$. Then we have

$$\mathbf{P} = -\frac{i}{\gamma\hbar} N d_{BX}^2 (\langle\cos^2\theta_x\rangle E_x \mathbf{e}_x + \langle\cos^2\theta_y\rangle E_y \mathbf{e}_y). \quad (4)$$

Here, $\langle\cos^2\theta_y\rangle$ and $\langle\cos^2\theta_x\rangle$ characterize the average alignment degrees of $N_2^+$ that are parallel and perpendicular to the polarization of the pump light, respectively, and $\langle\cos^2\theta_y\rangle$ is greater than $\langle\cos^2\theta_x\rangle$. Eq. (4) describes an anisotropic gain medium that has the minimum and maximum gian in the x and y directions, respectively.

Under the condition of a slowly varying approximation, the wave equation is given by

$$\frac{\partial \mathbf{E}}{\partial z} = \frac{i\omega}{2\varepsilon_0 c} \mathbf{P}, \quad (5)$$

where $\omega$ is the angular frequency of the electric field, $\varepsilon_0$ is the dielectric permittivity of vacuum, and *c* is light speed. Assuming the amplitude of the seed light's electric field is $E_0$, we can solve Eq. (5) and obtain the amplified electric field, namely,

$$\mathbf{E} = E_0 \sin\alpha e^{\langle\cos^2\theta_x\rangle Az} \mathbf{e}_x + E_0 \cos\alpha e^{\langle\cos^2\theta_y\rangle Az} \mathbf{e}_y, \quad (6)$$

where $A = N\omega d_{BX}^2 / 2\gamma\hbar\varepsilon_0 c$, characterizing the amplification coefficient. The total amplification factor of the seed light can be described by

$$g = \frac{|\mathbf{E}|^2}{|E_0|^2} = \sin^2\alpha e^{2\langle\cos^2\theta_x\rangle Az} + \cos^2\alpha e^{2\langle\cos^2\theta_y\rangle Az}. \quad (7)$$

The angle between the polarization directions of the SF (output electric field) and the pump light, $\alpha'$, satisfies

$$\sin\alpha' = \frac{E_x}{|\mathbf{E}|} = \frac{\sin\alpha}{\sqrt{1+\cos^2\alpha(e^{2(\langle\cos^2\theta_y\rangle-\langle\cos^2\theta_x\rangle)Az}-1)}}. \tag{8}$$

Eq. (7) and (8) constitute a parametric equation with $Az$ as the parameter, which can be utilized to characterize the dependence of $\alpha'$ on $g$. As shown in Fig. 3(a), the simulated curves are in good agreement with the experimental results, where the value of $\langle\cos^2\theta_y\rangle/\langle\cos^2\theta_x\rangle$ is fitted to 1.63.

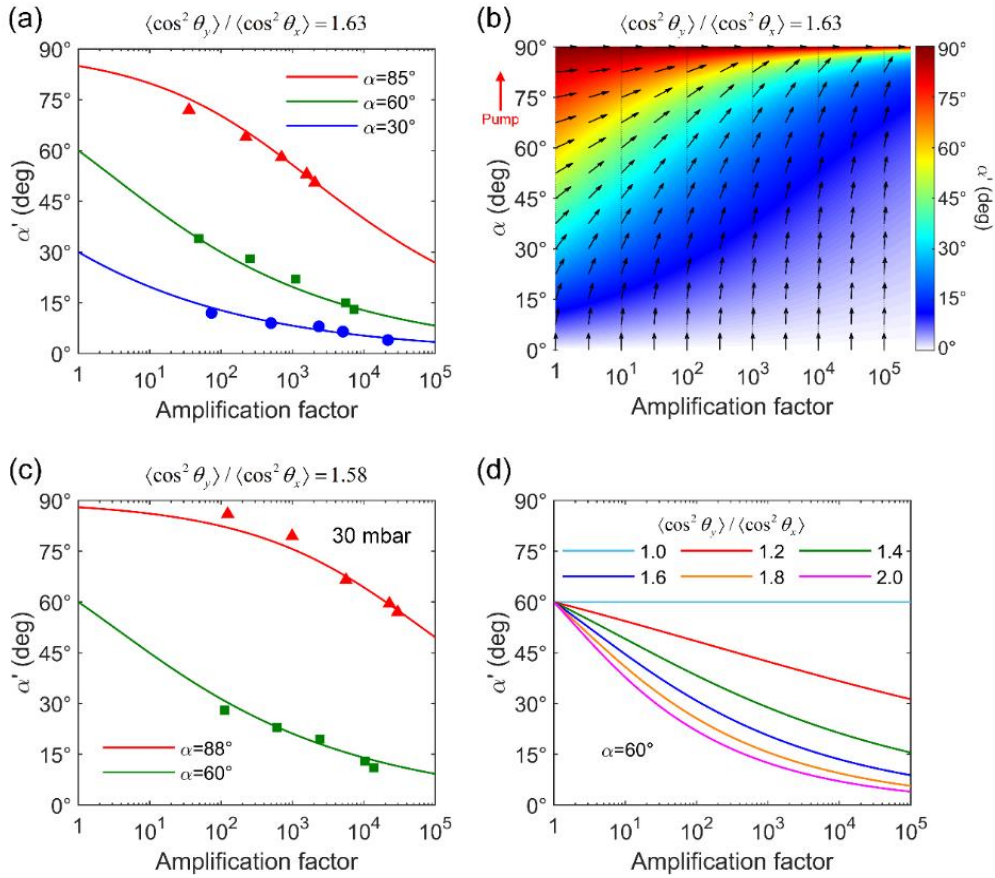

Fig. 3 (a) Measured and simulated $\alpha'$ versus $g$ with different $\alpha$ under 10 mbar. Dots and curves represent measured and simulated results, respectively. (b) Simulated $\alpha'$ versus $g$ and $\alpha$. The Red arrow represents the polarization direction of the pump light. Black arrows represent the polarization directions of the amplified signal at the location. (c) Measured and simulated $\alpha'$ versus $g$ with different $\alpha$ under 30 mbar. (d) Simulated $\alpha'$ versus $g$ with different alignment degrees when $\alpha = 60°$.

In Fig. 3(b), we comprehensively show simulated $\alpha'$ as a function of $\alpha$ and $g$. When the amplification factor is 1, the black arrows represent the polarization directions of the seed light at different angles. If $\alpha = 0°$ or $90°$, Eq. (8) will be simplified as $\sin\alpha' = \sin\alpha$. This interprets the previous observations that when the polarization of the seed light is parallel or perpendicular to that of the pump light, $\alpha'$ is always equal to $\alpha$ [25,26,35]. Under these conditions, the electric field of the injected seed light has only the $x$ or $y$ component, so after amplification, it maintains the $x$ or $y$ component. When $0° < \alpha < 90°$, $\alpha' < \alpha$ because the $y$-direction amplification factor is greater than the $x$-direction one. This explains why the polarization direction of the SF lies between those of the pump and seed light. Analyzing the monotonicity of Eq. (7) and (8), we can find that $g$ and $\alpha'$ monotonically increase and decrease with $Az$, respectively. As a result, $\alpha'$ is a monotonically decreasing function of $g$, as numerically demonstrated by the simulation results in Fig. 3(a) and (b). This explains the observation that the higher the amplification factor, the closer the polarization direction of the amplified signal is to that of the pump light. When the gas pressure is changed to 30 mbar, the experimental and simulated results are also well fitted, as shown in Fig. 3(c). We set $\alpha$ to 60° and show the simulated curves of $\alpha'$ versus $g$ at different alignment degrees in Fig. 3(d). When $\langle \cos^2 \theta_x \rangle = \langle \cos^2 \theta_y \rangle$, Eq. (8) becomes $\sin \alpha' = \sin \alpha$, which indicates the polarization of the amplified signal is always identical to that of the seed light in an isotropic gain amplifier. As $\langle \cos^2 \theta_y \rangle / \langle \cos^2 \theta_x \rangle$ becomes larger, the gain exhibits higher anisotropy, and thus the polarization direction of the SF becomes closer to that of the pump light.

The polarization characteristics of the R and P branches are slightly different. Previous studies demonstrated that the polarization directions of the R and P branches are symmetric relative to that of the pump light due to the negative coherence between $J$ and $J+2$ of $X^2\Sigma_g^+(v=0)$ [26,36]. Therefore, in the model, the $x$ component of the R-branch electric field is expressed as $E_x = E_0 \sin \alpha e^{\langle \cos^2 \theta_x \rangle Az} e^{-i\pi}$. In the experiment, the polarizations at the center and two sides of the R-branch spectrum are different, where the two sides are identical. As shown in Fig. 4, since the R-branch band is broad, the amplification factor varies significantly with the wavelength on a wide scale. According to the theoretical model, the right and left sides, connected by the dash lines, have the same amplification factor, leading to identical polarization directions. The polarization direction at the center differs from those of the two sides due to a larger amplification factor. As a final remark, we would like to point out that this work directly proves the assumption raised by our recent work [4], where a cylindrical vector pump beam induces a cylindrically symmetric anisotropic gain medium that amplifies a Gaussian seed beam into a cylindrical vector beam.

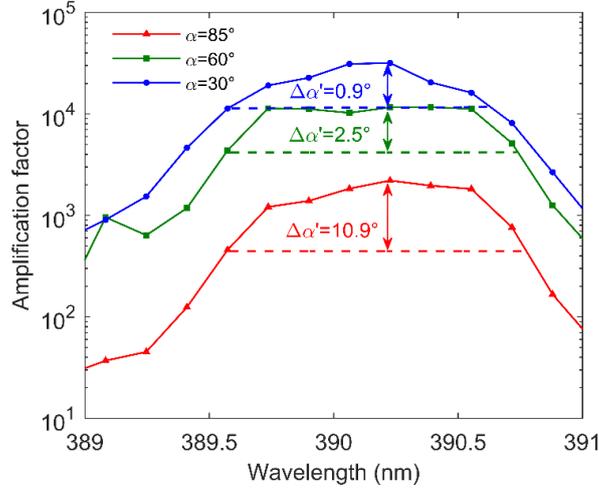

Fig. 4 Measured amplification factor versus the wavelength in the R-branch region. $\Delta\alpha'$ represents the angle difference of the polarization direction between the center and two sides.

In conclusion, we revealed the underlying mechanism of the polarization characteristics of the $N_2^+$ air lasing that has never been well understood for the past decade. We experimentally demonstrated that the optical gain is anisotropic, which is induced by the permanent alignment of $N_2^+$ due to the preferential ionization of the pump light. Owing to the anisotropy of the optical gain, as the amplification factor of the seed pulse increases, the polarization direction of the amplified SF progressively approaches that of the pump pulse. For the R branch, the nonuniformity of the amplification of the seed light at different wavelengths is responsible for their distinction of the polarization characteristic. We built a theoretical model and reproduced the experimental results well. According to the model, if the amplification factor, the polarization directions of the pump and the seed light, and the average alignment degree are confirmed, one can certainly predict the polarization angle of the amplified SF. This work gives the full interpretation of the polarization behaviors, which clarifies one fundamental puzzle and provides the basis for the comprehensive control of nitrogen ion air lasing in all dimensions.

## Acknowledgments

This work is supported by the National Key R&D Program of China (Grant No. 2018YFB2200401), and the National Natural Science Foundation of China (Grants Nos. 12034013, 12174011).

# References

[1] A. Dogariu, J. B. Michael, M. O. Scully, and R. B. Miles, *High-gain backward lasing in air*, Science **331**, 442-445 (2011).

[2] J. Gao, P. Zuo, T. Zhu, Q. Gong, and H. Jiang, *Study of the formation dynamics of OH from the photolysis of O3 by ultrashort laser pulses*, J. Phys. Chem. Lett. **11**, 6482-6486 (2020).

[3] H. Lei, J. Yao, J. Zhao, H. Xie, F. Zhang, H. Zhang, N. Zhang, G. Li, Q. Zhang, X. Wang, Y. Yang, L. Yuan, Y. Cheng, and Z. Zhao, *Ultraviolet supercontinuum generation driven by ionic coherence in a strong laser field*, Nat. Commun. **13**, 4080 (2022).

[4] J. Gao, X. Zhang, Y. Wang, Y. Fang, Q. Lu, Z. Li, Y. Liu, C. Wu, Q. Gong, Y. Liu, and H. Jiang, *Structured air lasing of $N_2^+$*, Commun. Phys. **6**, 97 (2023).

[5] H. Mei, J. Gao, K. Wang, J. Dong, Q. Gong, C. Wu, Y. Liu, H. Jiang, and Y. Liu, *Amplification of light pulses with orbital angular momentum (OAM) in nitrogen ions lasing*, Opt. Express **31**, 31912-31921 (2023).

[6] Y. Liu, P. Ding, G. Lambert, A. Houard, V. Tikhonchuk, and A. Mysyrowicz, *Recollision-induced superradiance of ionized nitrogen molecules*, Phys. Rev. Lett. **115**, 133203 (2015).

[7] Y. Liu, Y. Brelet, G. Point, A. Houard, and A. Mysyrowicz, *Self-seeded lasing in ionized air pumped by 800 nm femtosecond laser pulses*, Opt. Express **21**, 22791-22798 (2013).

[8] H. Xu, E. Lötstedt, A. Iwasaki, and K. Yamanouchi, *Sub-10-fs population inversion in $N_2^+$ in air lasing through multiple state coupling*, Nat. Commun. **6**, 8347 (2015).

[9] J. Yao, S. Jiang, W. Chu, B. Zeng, C. Wu, R. Lu, Z. Li, H. Xie, G. Li, C. Yu, Z. Wang, H. Jiang, Q. Gong, and Y. Cheng, *Population redistribution among multiple electronic states of molecular nitrogen ions in strong laser fields*, Phys. Rev. Lett. **116**, 143007 (2016).

[10] Y. Liu, P. Ding, N. Ibrakovic, S. Bengtsson, S. Chen, R. Danylo, E. R. Simpson, E. W. Larsen, X. Zhang, and Z. Fan, *Unexpected sensitivity of nitrogen ions superradiant emission on pump laser wavelength and duration*, Phys. Rev. Lett. **119**, 203205 (2017).

[11] M. Lei, C. Wu, A. Zhang, Q. Gong, and H. Jiang, *Population inversion in the rotational levels of the superradiant $N_2^+$ pumped by femtosecond laser pulses*, Opt. Express **25**, 4535-4541 (2017).

[12] M. Britton, P. Laferrière, D. H. Ko, Z. Li, F. Kong, G. Brown, A. Naumov, C. Zhang, L. Arissian, and P. B. Corkum, *Testing the role of recollision in $N_2^+$ air lasing*, Phys. Rev. Lett. **120**, 133208 (2018).

[13] A. Zhang, M. Lei, J. Gao, C. Wu, Q. Gong, and H. Jiang, *Subfemtosecond-resolved modulation of superfluorescence from ionized nitrogen molecules by 800-nm femtosecond laser pulses*, Opt. Express **27**, 14922-14930 (2019).

[14] A. Zhang, Q. Liang, M. Lei, L. Yuan, Y. Liu, Z. Fan, X. Zhang, S. Zhuang, C. Wu, Q. Gong, and H. Jiang, *Coherent modulation of superradiance from nitrogen ions pumped with femtosecond pulses*, Opt. Express **27**, 12638-12646 (2019).

[15] J. Chen, J. Yao, H. Zhang, Z. Liu, B. Xu, W. Chu, L. Qiao, Z. Wang, J. Fatome, O. Faucher, C. Wu, and Y. Cheng, *Electronic-coherence-mediated molecular nitrogen-ion lasing in a strong laser field*, Phys. Rev. A **100**, 031402(R) (2019).

[16] J. Yao, W. Chu, Z. Liu, B. Xu, J. Chen, and Y. Cheng, *Generation of Raman lasers from nitrogen molecular ions driven by ultraintense laser fields*, New J. Phys. **20**, 033035 (2018).

[17] Z. Liu, J. Yao, J. Chen, B. Xu, W. Chu, and Y. Cheng, *Near-resonant Raman amplification in the rotational quantum wave packets of nitrogen molecular ions generated by strong field ionization*, Phys. Rev. Lett. **120**, 083205 (2018).

[18] M. Richter, M. Lytova, F. Morales, S. Haessler, O. Smirnova, M. Spanner, and M. Ivanov, *Rotational quantum beat lasing without inversion*, Optica **7**, 586-592 (2020).

[19] X. Zhang, Q. Lu, Z. Zhang, Z. Fan, D. Zhou, Q. Liang, L. Yuan, S. Zhuang, K. Dorfman, and Y. Liu, *Coherent control of the multiple wavelength lasing of $N_2^+$: coherence transfer and beyond*, Optica **8**, 668-673 (2021).

[20] J. Yao, B. Zeng, H. Xu, G. Li, W. Chu, J. Ni, H. Zhang, S. L. Chin, Y. Cheng, and Z. Xu, *High-brightness switchable multiwavelength remote laser in air*, Phys. Rev. A **84**, 051802(R) (2011).

[21] A. Azarm, P. Corkum, and P. Polynkin, *Optical gain in rotationally excited nitrogen molecular ions*, Phys. Rev. A **96**, 051401(R) (2017).

[22] A. Mysyrowicz, R. Danylo, A. Houard, V. Tikhonchuk, X. Zhang, Z. Fan, Q. Liang, S. Zhuang, L. Yuan, and Y. Liu, *Lasing without population inversion in $N_2^+$*, APL Photonics, **4**, 110807 (2019)

[23] C. Kleine, M.O. Winghart, Z.Y. Zhang, M. Richter, M. Ekimova, S. Eckert, M. J. J. Vrakking, E. T. J. Nibbering, A. Rouzée, and E. R. Grant, *Electronic state population dynamics upon ultrafast strong field ionization and fragmentation of molecular nitrogen*, Phys. Rev. Lett. **129**, 123002 (2022).

[24] Q. Zhang, J. Zhao, G. Bai, B. Zhang, W. Tao, Q. Qiu, H. Lei, Y. Lang, J. Liu, X. Wang, Z. Zhao, *Probing electronic-vibrational dynamics of N2+ induced by strong-field ionization*, arXiv preprint arXiv:2310.04210 (2023).

[25] J. Yao, G. Li, C. Jing, B. Zeng, W. Chu, J. Ni, H. Zhang, H. Xie, C. Zhang, and H. Li, *Remote creation of coherent emissions in air with two-color ultrafast laser pulses*, New J. Phys. **15**, 023046 (2013).

[26] H. Xie, H. Lei, G. Li, Q. Zhang, X. Wang, J. Zhao, Z. Chen, J. Yao, Y. Cheng, and Z. Zhao, *Role of rotational coherence in femtosecond-pulse-driven nitrogen ion lasing*, Phys. Rev. Research **2**, 023329 (2020).

[27] A. Forbes, M. de Oliveira, and M. R. Dennis, *Structured light*, Nat. Photon. **15**, 253–262 (2021).

[28] C. He, Y. Shen, and A. Forbes, *Towards higher-dimensional structured light*, Light Sci. Appl. **11**, 205 (2022).

[29] Y. Shen, Q. Zhang, P. Shi, L. Du, A. V. Zayats, and X. Yuan, *Topological quasiparticles of light: optical skyrmions and beyond*, arXiv preprint, arXiv:2205.10329 (2022).

[30] D. Sugic, R. Droop, E. Otte, D. Ehrmanntraut, F. Nori, J. Ruostekoski, C. Denz, and M. R. Dennis, *Particle-like topologies in light*, Nat. Commun. **12**, 6785 (2021)

[31] A. de las Heras, A. K. Pandey, J. San Román, J. Serrano, E. Baynard, G. Dovillaire, M. Pittman, C. G. Durfee, L. Plaja, S. Kazamias, O. Guilbaud, and C. Hernández-García, *Extreme-ultraviolet vector-vortex beams from high harmonic generation*, Optica **9**, 71-79 (2022)

[32] M. Lei, C. Wu, Q. Liang, A. Zhang, Y. Li, Q. Cheng, S. Wang, H. Yang, Q. Gong, and H. Jiang, *The fast decay of ionized nitrogen molecules in laser filamentation investigated by a picosecond streak camera*, J. Phys. B: At. Mol. Opt. Phys. **50**, 145101 (2017).

[33] S. Petretti, Y. V. Vanne, A. Saenz, A. Castro, and P. Decleva, *Alignment-dependent ionization of $N_2$, $O_2$, and $CO_2$ in intense laser fields*, Phys. Rev. Lett. **104**, 223001 (2010).

[34] M. O. Scully and M. S. Zubairy, *Quantum Optics* (Cambridge University Press, Cambridge ; New York, 1997).

[35] J. Ni, W. Chu, C. Jing, H. Zhang, B. Zeng, J. Yao, G. Li, H. Xie, C. Zhang, H. Xu, S. Chin, Y. Cheng, Z. Xu, *Identification of the physical mechanism of generation of coherent $N_2^+$ emissions in air by femtosecond laser excitation*, Opt. Express **21**, 8746-8752 (2013).

[36] H. Lei, G. Li, H. Xie, Q. Zhang, X. Wang, J. Zhao, Z. Chen, and Z. Zhao, *Mechanism and control of rotational coherence in femtosecond laser-driven $N_2^+$*, Opt. Express **28**, 22829-22843 (2020).